\documentclass[conference]{IEEEtran}

\usepackage{cite}
\usepackage{amsmath,amssymb}
\usepackage{graphicx}
\usepackage{booktabs}
\usepackage{multirow}
\usepackage{url}
\usepackage{stfloats}

\title{Quantum Computing for Network Security Classification: Near-Term Classification and Long-Term Memory Efficiency}

\author{
\IEEEauthorblockN{
Yuqing Li$^{1}$,
Poonam Bala Nehru$^{2}$,
Yunpeng Zhang$^{2}$,
Danindu Gammanpilage$^{2}$,
\\
Xin Jin$^{1}$,
Zeguan Wu$^{1}$,
and Junyu Liu$^{1,*}$
}
\IEEEauthorblockA{$^{1}$University of Pittsburgh, Pittsburgh, PA, USA\\
Emails: yul658@pitt.edu, xij90@pitt.edu, zew79@pitt.edu,\\
junyuliu@pitt.edu}
\IEEEauthorblockA{$^{2}$University of Houston, Houston, TX, USA\\
Emails: pbalaneh@cougarnet.uh.edu, yzhan226@central.uh.edu, dgammanp@cougarnet.uh.edu}
\IEEEauthorblockA{$^{*}$Corresponding author: Junyu Liu (junyuliu@pitt.edu)}
}

\begin{document}
\maketitle

\begin{abstract}
As an emerging technology, quantum computing has already been explored in
several network-security applications. However, how quantum computing may
contribute to network-security classification in both the near term and the
longer term has not been systematically discussed. This paper studies this
question through two complementary experiments. First, we evaluate near-term
quantum-kernel support vector machines (SVMs) on practical network-security
classification tasks and compare them with classical SVM baselines on
KDD Cup 1999, CICIDS2017, and BoT-IoT. Across these runs, quantum kernels are
competitive but task-dependent: they can match or improve classical baselines in
some settings, while classical RBF kernels remain
stronger in others. This suggests that near-term quantum-kernel methods should
be evaluated as practical, dataset-dependent alternatives to classical kernels
rather than as uniformly superior replacements. Second, we use quantum oracle
sketching (QOS) to study a longer-term memory advantage for classification with
streaming classical samples. In QOS, samples
are processed online and used to incrementally construct an approximate quantum
oracle, which provides coherent query access for downstream quantum algorithms
without retaining the entire dataset. Under the QOS-inspired machine-size
estimate, comparable accuracy corresponds to a substantially smaller effective
memory-size proxy than explicit sparse/QRAM-style storage. Compared with a
simple streaming proxy, the result is more nuanced because aggressive feature
filtering can make the streaming dimension very small. This suggests that the
long-term value of quantum computing for network-security classification may lie
in memory-efficient data access rather than immediate runtime speedup.
Together, these experiments show how quantum computing may contribute to
network-security classification from two perspectives: near-term classification
performance and longer-term memory efficiency.
\end{abstract}


\section{Introduction}

Quantum computing has recently been applied to network-security classification.
In this paper, we use network-security classification broadly to include
supervised binary classification and one-class anomaly detection with binary
decisions. Existing studies have reported high accuracy on intrusion detection,
DDoS detection, anomaly detection, and attack-type classification tasks,
suggesting that quantum and hybrid quantum-classical models have potential for this domain.
Prior work has explored
quantum-assisted intrusion detection with classical preprocessing and quantum
simulation~\cite{gouveia2020quantum_nids}, quantum support vector machines and
hybrid quantum-classical neural networks for DDoS detection~\cite{payares2021qml_ddos},
quantum neural networks for network anomaly detection on noisy quantum
computers~\cite{kukliansky2024qnn_anomaly}, and broader QML-based systems for intrusion
classification evaluated on binary and multiclass security tasks~\cite{abreu2024qml_ids,
	abreu2025quantumnetsec}. These results indicate that quantum models can be
competitive for selected network-security classification problems, while also
showing that performance depends strongly on the dataset, model architecture,
and evaluation setting.

However, most existing studies evaluate quantum methods mainly from the
perspective of near-term classification accuracy: whether a specific quantum or
hybrid model improves performance on a specific
dataset. This motivates a broader framing of how quantum computing may
contribute to network-security classification across different time scales. In the near term,
quantum models should be compared with classical baselines under controlled
datasets, preprocessing, and evaluation metrics. In the longer term, quantum
methods may contribute not only through accuracy, but also through lower
memory-size requirements for processing large-scale classical security data.

This paper studies network-security classification from these two perspectives.
In the near-term experiment, we use two binary SVM settings: one-class anomaly
scoring for normal-versus-attack decisions, and supervised binary
classification. A one-class anomaly method is binary at the decision-output
level because a test sample is classified as either normal/benign or
anomalous/attack, but it is not a supervised binary classifier at the
training-objective level. We therefore treat it as a special case of binary
network-security classification, which also keeps the experimental scope consistent with the binary
classification setting considered in the QOS-inspired memory analysis. Unlike
prior work that often proposes a specific quantum or hybrid IDS model and
reports its accuracy, our near-term experiment focuses on a controlled
kernel-level comparison. Quantum and classical SVMs use the same sampled data,
preprocessing, and reduced angle features, differing mainly in the
similarity function used by the SVM. This design isolates whether the
quantum fidelity kernel itself provides a competitive classification signal
when the data pipeline and SVM objective are kept fixed, rather than attributing performance differences to
different preprocessing pipelines or model families. Specifically, we evaluate
quantum-kernel support vector machines (SVMs) on binary network-security
classification tasks derived from KDD Cup 1999, CICIDS2017, and
BoT-IoT, and compare
them with classical SVM baselines. This experiment tests whether quantum
fidelity kernels are competitive with classical kernels when the classification
setting is fixed. Our results show a dataset-dependent pattern: quantum kernels can match or
improve classical baselines in some settings, while classical RBF kernels remain
stronger in others. This is consistent with prior network-security QML studies,
which report strong quantum or hybrid performance on selected datasets but also
show sensitivity to dataset characteristics, model design, and evaluation
settings. Thus, the near-term value of quantum-kernel
SVMs appears to be dataset-dependent rather than universally superior. The main
near-term contribution is therefore not a claim of clear or statistically
uniform accuracy advantage. Instead, it is a controlled comparison that shows
where quantum kernels are competitive under the same SVM objective and
preprocessing pipeline, and where classical kernels remain stronger.

The first experiment addresses near-term classification performance. The second part is motivated by a
longer-term result: quantum oracle sketching (QOS)~\cite{zhao2026qos} showed an
exponential machine-size advantage for binary classification on massive
classical data. We use this result as a lens for network-security
classification. Instead of assuming that the full data matrix must be kept in
memory, QOS processes classical samples one at a time and builds an approximate
quantum oracle for later computation. Following this machine-size estimate,
we evaluate where this memory-size advantage appears on network-security
classification tasks by checking whether comparable accuracy can be maintained
with a smaller effective memory-size proxy. Our results show that the
QOS-based estimate is much smaller than explicit sparse-storage or QRAM-style
storage at comparable accuracy levels. Compared with the classical streaming
proxy, the advantage is more nuanced: streaming can be smaller after aggressive
feature filtering, while QOS models compact coherent data access rather than
only the dimension of a fitted streaming classifier.

In summary, this paper organizes quantum computing for network-security
classification around two parts. The first part evaluates quantum-kernel SVMs
against classical SVM baselines on practical binary network-security tasks and
uses the mixed results to characterize quantum kernels as task-dependent
alternatives rather than universal replacements. The second part uses the QOS
machine-size estimate to evaluate a longer-term memory advantage for analogous
binary tasks. We study these
parts on KDD Cup 1999,
CICIDS2017, and BoT-IoT, using matched preprocessing and evaluation metrics
where applicable. The detailed dataset setup, model construction, and evaluation
protocol are described in the next section.

\section{Experimental Design}

\subsection{Datasets and Binary Tasks}

We evaluate both experimental parts on three network-security datasets:
KDD Cup 1999~\cite{kdd1999}, CICIDS2017~\cite{cicids2017}, and
BoT-IoT~\cite{botiot}. The goal is not to cover every possible
attack category, but to build controlled binary network-security
tasks that can be used consistently across the near-term SVM comparison and the
longer-term QOS memory analysis.

For KDD Cup 1999 and CICIDS2017, we use a one-class anomaly setting.
The model is trained only on normal or benign samples and evaluated on a test
set containing both normal/benign and attack samples. Although this one-class
setting is not trained as a standard supervised binary classifier, its output
decision is binary: each test sample is labeled as normal/benign or
anomalous/attack. Thus, it is binary at the decision-output level but not at the
training-objective level. For BoT-IoT, the dataset subset used in our
experiments contains too few normal-flow samples for stable one-class training.
We therefore use a supervised DoS-versus-DDoS binary task and report it
separately from the one-class anomaly tasks.

\begin{figure*}[t]
	\centering
	\includegraphics[width=0.98\textwidth]{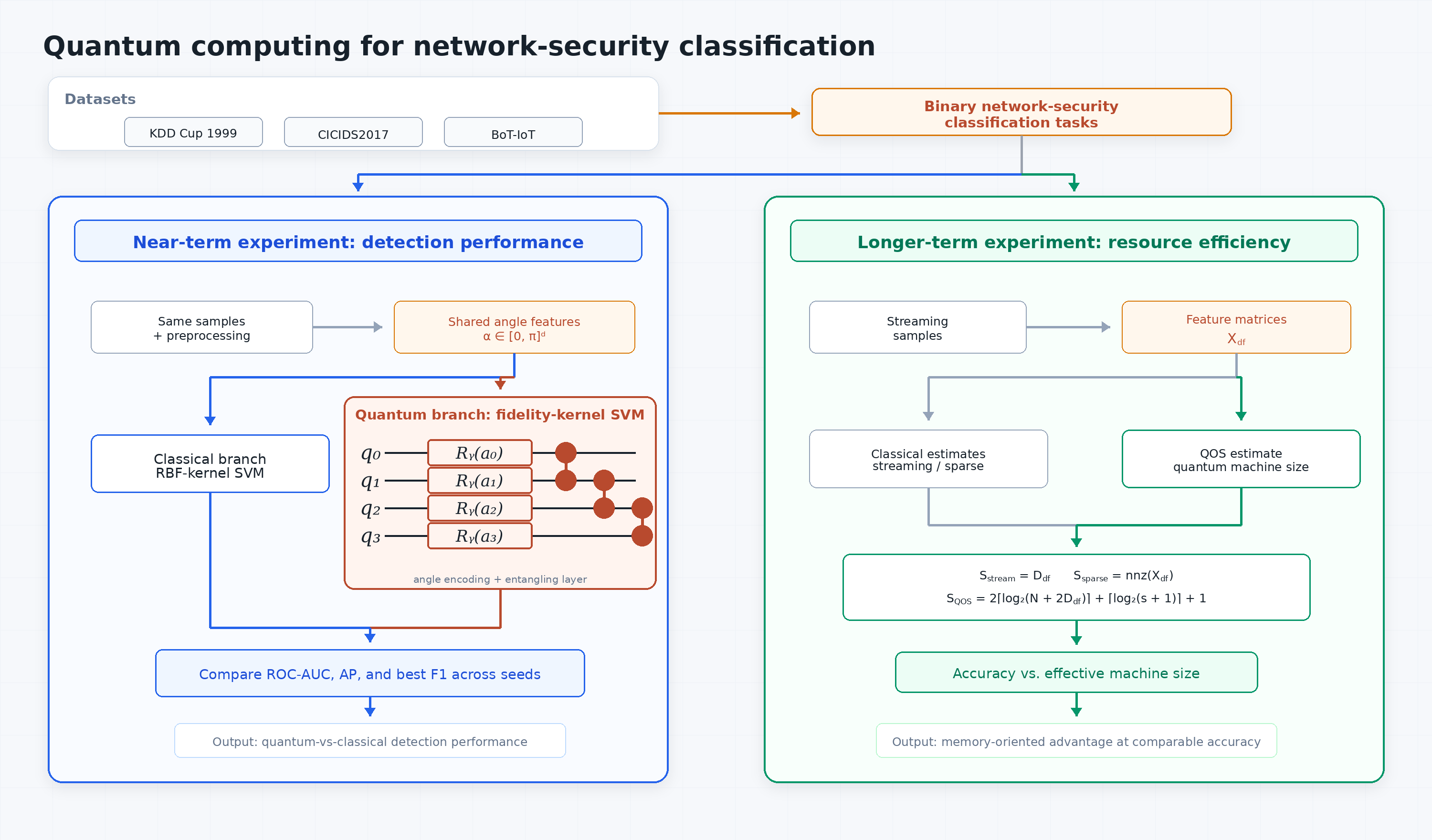}
	\caption{Overall workflow for the two experiments. The near-term branch
	compares classical and quantum SVM kernels for network-security detection and
	classification performance, while the longer-term branch evaluates
	QOS-inspired resource efficiency through effective machine-size estimates.}
	\label{fig:overall-workflow}
\end{figure*}

Fig.~\ref{fig:overall-workflow} summarizes the full experimental workflow. The
three datasets are first converted into binary network-security classification
tasks. The left branch evaluates near-term detection and classification
performance by comparing classical RBF-kernel SVMs with quantum fidelity-kernel
SVMs under the same samples, preprocessing, and angle features. The right branch
evaluates longer-term resource efficiency by comparing classical
streaming/sparse machine-size estimates with the QOS-inspired quantum
machine-size estimate at comparable accuracy.

\subsection{Near-Term SVM Comparison}

In the near-term experiment, we compare quantum-kernel SVMs with classical SVM
baselines using the same sampled rows, the same train/test split, and the same
low-dimensional angle features. The preprocessing pipeline is fit only on
the training data and then applied to the test data.

For KDD Cup 1999 and CICIDS2017, the training set contains only normal or
benign samples, while the test set contains both normal/benign and attack
samples. The one-class SVM therefore learns the normal region and assigns an
anomaly score to each test sample. For BoT-IoT, because the dataset subset used
in our experiments contains too few normal flows for stable one-class training,
we use a supervised DoS-versus-DDoS task with balanced samples from the two
attack classes.

After sampling, we remove label and identifier columns, one-hot encode
categorical variables, and standardize numerical variables using the training
set. The preprocessing and SVM baselines use standard scikit-learn
routines~\cite{pedregosa2011scikit}. This gives a sparse numerical feature
matrix for each dataset. We then use TruncatedSVD~\cite{halko2011finding} to
map the encoded feature matrix to a shared
$d$-dimensional feature space, where $d$ is set equal to the number of qubits
used by the quantum kernel. Specifically, TruncatedSVD is fit on the encoded training matrix and
applied to both train and test matrices:
\begin{equation}
	Z_{\mathrm{train}} = X_{\mathrm{train}} V_d,\qquad
	Z_{\mathrm{test}} = X_{\mathrm{test}} V_d,
\end{equation}
where $d=n_{\mathrm{qubits}}$. The SVD coordinates are standardized using the
training set statistics. They are then scaled to angle features in $[0,\pi]$:
\begin{equation}
	A_{ij}=\pi\cdot
	\frac{Z_{ij}-\min_j(Z_{\mathrm{train}})}
	{\max_j(Z_{\mathrm{train}})-\min_j(Z_{\mathrm{train}})}.
\end{equation}
The same fitted scaler is applied to the test set, and test angles are clipped
to $[0,\pi]$. In the main runs, KDD Cup 1999 and CICIDS2017 use
$d=12$, while BoT-IoT uses $d=10$.

Both classical and quantum SVMs use these same angle-scaled SVD features as
input. The classical baseline applies an RBF kernel to the angle features. The
quantum model instead encodes each angle vector $a$ into a quantum feature state
$|\phi(a)\rangle$ and uses the fidelity
kernel~\cite{havlicek2019supervised,schuld2019quantum_feature}
\begin{equation}
	K(a,b)=|\langle\phi(a)|\phi(b)\rangle|^2.
\end{equation}
In the main experiment, we use the entangled feature map implemented in the
code, which applies single-qubit $R_y$ rotations followed by nearest-neighbor
phase interactions. The resulting kernel matrix is computed by exact
statevector simulation in row blocks to control memory.

For KDD Cup 1999 and CICIDS2017, both the classical and quantum models use the
one-class SVM objective~\cite{scholkopf2001estimating}. For BoT-IoT, both use
the supervised SVC objective~\cite{cortes1995support}.
Thus, within each dataset, the main difference between the classical and quantum
models is the kernel function rather than the sampled data, preprocessing, or
SVM objective. Because KDD Cup 1999/CICIDS2017 and BoT-IoT use different
training objectives, the comparisons should be interpreted within each dataset
and objective setting rather than as direct cross-dataset comparisons of
identical learning problems. The main comparison reported in
Table~\ref{tab:main-results} uses eight random seeds. We report ROC-AUC,
average precision (AP), and best F1 as mean $\pm$ standard deviation across
these seeds. Best F1 is computed by sweeping the decision threshold on the
evaluation scores; it should therefore be interpreted as a threshold-optimized
summary rather than as a fixed-threshold deployment metric.

\subsection{Longer-Term QOS Memory-Advantage Analysis}

The second experiment studies a longer-term memory advantage motivated by
quantum oracle sketching (QOS)~\cite{zhao2026qos}. The QOS paper considers
large-scale binary classification on massive classical data and shows that,
under sparsity and conditioning assumptions, a small quantum machine can achieve
the same prediction task with exponentially smaller machine size than classical
machines. Here, we use machine size as an effective memory-size proxy rather
than as a runtime measure. In particular, for a training matrix
$X\in\mathbb{R}^{N\times D}$ and binary labels
$y\in\{+1,-1\}^N$, the classification rule is based on the ridge or LS-SVM
weight vector
\begin{equation}
	w=(X^\top X+\lambda I)^{-1}X^\top y,
\end{equation}
and a test vector $x'$ is classified by
\begin{equation}
	\hat{y}=\mathrm{sign}(x'^\top w).
\end{equation}

The key idea is not that the full dataset is stored in a quantum state. Rather,
QOS treats the data as samples observed one at a time. These samples are used to
construct an approximate quantum oracle for data access, which
can be called by downstream quantum algorithms. Under the assumptions of the QOS
framework, this enables binary classification with quantum machine size
$\mathrm{poly}(\log D)$, while classical machines below a much larger size may
require super-polynomially more samples. In this sense, QOS changes the memory
target from storing the full data matrix to building a much smaller machine that
can still maintain the classification rule.

To connect this idea to network-security classification, we follow the
machine-size advantage perspective of QOS. For each dataset, we vary a feature
filtering threshold, denoted $\mathrm{min\_df}$, which removes features that
appear in too few samples. This produces a sequence of feature matrices
$X_{\mathrm{df}}\in\mathbb{R}^{N\times D_{\mathrm{df}}}$ with different
effective dimensions and sparsity levels. For each matrix, we train a ridge
classifier and record the classification accuracy.

For the same feature matrix, we then estimate three machine sizes:
\begin{align}
	S_{\mathrm{stream}} &= D_{\mathrm{df}},\\
	S_{\mathrm{sparse}} &= \mathrm{nnz}(X_{\mathrm{df}}),\\
	S_{\mathrm{QOS}} &=
	2\left\lceil \log_2(N+2D_{\mathrm{df}})\right\rceil
	+\left\lceil \log_2(s+1)\right\rceil + 1,
\end{align}
where $\mathrm{nnz}(X_{\mathrm{df}})$ is the number of nonzero entries and
$s$ is the maximum row or column sparsity. The first estimate represents a
classical streaming baseline, the second represents explicit sparse storage or
QRAM-style storage, and the third is the QOS-inspired quantum machine-size
estimate used in our experiment.

This experiment is therefore an estimate-based memory-size analysis, not a
runtime or hardware speedup claim. Its purpose is to separate two comparisons:
QOS versus explicit sparse/QRAM-style storage, where the QOS-inspired estimate
is much smaller, and QOS versus a simple streaming proxy, where the result
depends on how aggressively the feature space is filtered.

\section{Near-Term SVM Results}

\begin{table*}[t]
\centering
\caption{Main binary SVM results. Values are mean $\pm$ standard deviation
across eight random seeds.}
\label{tab:main-results}
\begin{tabular}{llccc}
\toprule
Dataset & Model & ROC-AUC & Average precision & Best F1 \\
\midrule
KDD Cup 1999 & Classical RBF OCSVM & \textbf{\boldmath$0.9282 \pm 0.0105$} & \textbf{\boldmath$0.9460 \pm 0.0086$} & \textbf{\boldmath$0.9092 \pm 0.0207$} \\
KDD Cup 1999 & Quantum OCSVM & $0.8972 \pm 0.0303$ & $0.9288 \pm 0.0195$ & $0.8891 \pm 0.0230$ \\
\midrule
CICIDS2017 & Classical RBF OCSVM & $0.6209 \pm 0.0073$ & $0.6975 \pm 0.0069$ & $0.6789 \pm 0.0012$ \\
CICIDS2017 & Quantum OCSVM & \textbf{\boldmath$0.6666 \pm 0.0316$} & \textbf{\boldmath$0.7034 \pm 0.0274$} & \textbf{\boldmath$0.7003 \pm 0.0060$} \\
\midrule
BoT-IoT & Classical RBF SVC & \textbf{\boldmath$0.9985 \pm 0.0002$} & \textbf{\boldmath$0.9985 \pm 0.0003$} & $0.9887 \pm 0.0021$ \\
BoT-IoT & Quantum SVC & $0.9977 \pm 0.0007$ & $0.9977 \pm 0.0005$ & \textbf{\boldmath$0.9955 \pm 0.0005$} \\
\bottomrule
\end{tabular}
\end{table*}

Table~\ref{tab:main-results} shows that the near-term result is mixed and
dataset-dependent. On KDD Cup 1999, the classical RBF OCSVM is stronger on all
three metrics, indicating that the RBF geometry is well matched to this
low-dimensional feature space. On CICIDS2017, the quantum OCSVM improves
ROC-AUC, AP, and best F1, suggesting that the quantum fidelity kernel captures
a useful similarity structure for BENIGN-versus-attack classification. On BoT-IoT,
both models are near saturation. The classical RBF SVC has slightly higher
ROC-AUC and AP, while the quantum SVC has consistently higher best F1.

This variation is likely related to the geometry of each task after the shared
SVD-to-angle preprocessing. On KDD Cup 1999, the reduced normal-versus-attack
space appears well matched to the distance-based RBF OCSVM, while the periodic
angle encoding used by the quantum kernel may distort this radial structure. On
CICIDS2017, benign and attack flows are more heterogeneous, so the nonlinear
interactions introduced by the fidelity kernel can provide a better ranking
signal. On BoT-IoT, both kernels are near saturation for the supervised
DoS-versus-DDoS task; the higher quantum best F1 mainly reflects a more favorable
threshold on this balanced test set rather than a broad ranking advantage.
Overall, the table supports controlled comparability and dataset-dependent
behavior, not a uniform near-term quantum advantage.

\begin{figure*}[!b]
\centering
\includegraphics[width=0.98\textwidth]{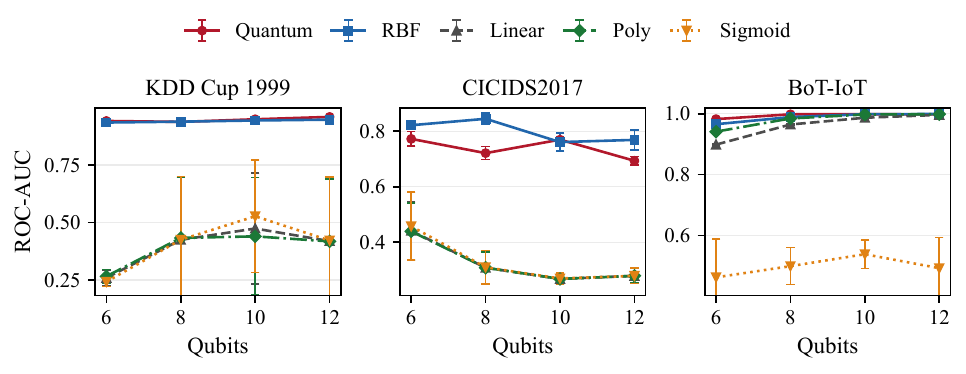}
\caption{ROC-AUC ablation across qubit counts and kernel choices. Error bars
show standard deviation across three seeds.}
\label{fig:auc-ablation}
\end{figure*}

We also test whether the comparison depends strongly on the classical kernel or
qubit count. For computational tractability, this auxiliary ablation uses three
seeds and smaller samples: 50k normal or benign training samples for
KDD/CICIDS2017 and 25k samples per class for BoT-IoT. We compare quantum
kernels with RBF, linear, polynomial, and sigmoid
SVM kernels across 6, 8, 10, and 12 qubits. Fig.~\ref{fig:auc-ablation} shows
the ROC-AUC trend. The corresponding AP and best-F1 ablations show the same
main pattern summarized below. RBF is the strongest classical
baseline among the tested classical kernels, so the main comparison is not
against a weak baseline. Increasing the number of qubits is not monotonic:
useful dimension depends on dataset and metric. In this smaller-sample
ablation, quantum kernels are strongest for BoT-IoT best F1 and sometimes
achieve the strongest KDD ROC-AUC/AP, even though the full main experiment in
Table~\ref{tab:main-results} favors the RBF OCSVM on KDD. This difference
suggests sensitivity to sampling scale and reinforces the dataset- and
setting-dependent nature of the comparison. These results support a claim of
controlled comparability rather than universal quantum advantage.

\section{Oracle Sketching and Memory Efficiency}

The near-term SVM experiment evaluates classification performance in our
simulation setting. The oracle-sketching experiment complements it with a memory
analysis: for the same datasets and corresponding binary labels, it compares
the effective memory-size proxies associated with classical storage baselines
and with the QOS-inspired compact-oracle estimate at comparable accuracy.

Following the feature-truncation strategy used in QOS, we vary a minimum
feature-frequency threshold, denoted $\mathrm{min\_df}$. A feature is kept only
if it is present in at least $\mathrm{min\_df}$ samples; increasing
$\mathrm{min\_df}$ removes rarer features and reduces the effective feature
dimension. For each threshold, we train a ridge classifier and record the
resulting accuracy. We then compare three machine-size estimates. The classical
streaming size is the feature dimension required by a streaming baseline. The
classical sparse/QRAM-style size counts explicit sparse storage. The quantum
oracle sketching size is the compact-oracle estimate, which scales
logarithmically with dimensions and sparsity in our implementation.

\begin{figure*}[t]
\centering
\includegraphics[width=0.98\textwidth]{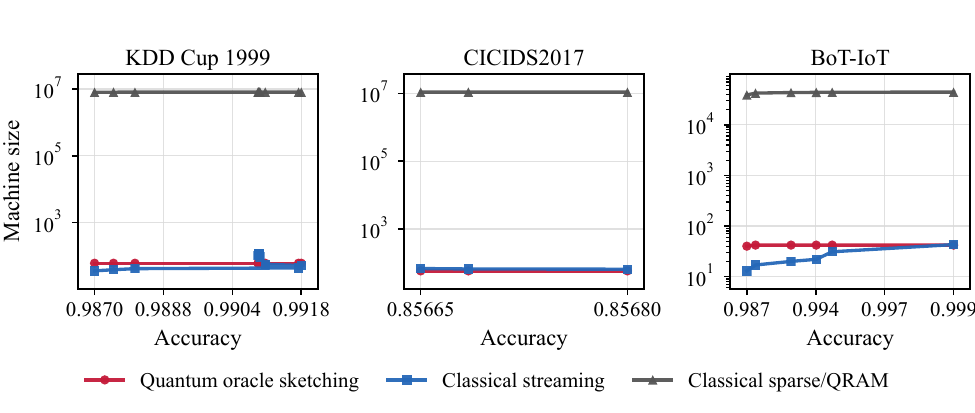}
\caption{Machine size versus accuracy for the quantum oracle sketching
experiment. Because the accuracies are near-saturated for some datasets, the
x-axis is zoomed to the observed accuracy range for readability.}
\label{fig:oracle-sketch}
\end{figure*}

Fig.~\ref{fig:oracle-sketch} shows the memory-oriented comparison. In the
figure, red denotes the QOS estimate, blue denotes the classical streaming
proxy, and gray denotes explicit sparse/QRAM-style storage. Across all three
datasets, comparable accuracy is associated with a much smaller effective
memory-size proxy than explicit sparse/QRAM-style storage. The comparison with
the streaming proxy is not uniformly in favor of QOS, because the streaming
estimate only counts the reduced feature dimension and can become very small
when rare features are aggressively removed. This result should not be
interpreted as a runtime speedup for the current simulator, and it does not
model the engineering cost of building a real quantum data-access mechanism.
Instead, it highlights a more specific longer-term point: even when near-term
quantum kernels are only comparable to classical kernels, compact quantum data
access may still reduce the memory burden relative to explicit data storage for
large network-security classification tasks.

\section{Discussion}

The two experiments address network-security classification from two time
scales. The SVM experiment evaluates whether quantum kernels can serve as
alternatives to classical kernels under the same preprocessing pipeline and SVM
objectives. The oracle-sketching experiment evaluates a longer-term
memory-size perspective, especially the gap between compact quantum-oracle
estimates and explicit sparse/QRAM-style storage. Interpreted together, they
suggest that quantum computing may support network-security classification in
more than one way: as task-dependent near-term kernel alternatives for
classification tasks, and as memory-efficient data-access mechanisms for future
large-scale classification tasks.

These two experiments should not be interpreted as a single end-to-end quantum
network-security system. The near-term experiment evaluates kernel-level
classification performance under simulation, whereas the QOS experiment provides
an estimate-based memory-size analysis motivated by future quantum data-access
models. Their common role is to identify two possible contribution channels for
quantum computing, not to demonstrate a unified deployed quantum IDS pipeline.

\section{Conclusion}

This paper studies how quantum computing may contribute to network-security
classification along two time scales: near-term classification performance and longer-term memory
efficiency. In the near term, quantum kernel SVMs are competitive but
task-dependent: they improve CICIDS2017 metrics, improve BoT-IoT best F1, and
underperform the RBF baseline on KDD. In the longer term, quantum oracle
sketching suggests a separate memory advantage relative to explicit
sparse/QRAM-style storage, while the comparison with a simple streaming proxy
depends on feature filtering. The central conclusion is therefore measured but
optimistic: quantum computing may support future network-security classification
both through near-term kernel alternatives for classification tasks and through
longer-term memory-efficient data access for large-scale classification.

\section*{Acknowledgement}
YL, XJ, ZW, and JL are supported in part by the University of Pittsburgh, School of Computing and Information, Department of Computer Science, Pitt Cyber, Pitt Momentum fund, PQI Community Collaboration Awards, John C. Mascaro Faculty Scholar in Sustainability, Switzerland NSF award 2000-1-243053, NSF award 2535915, 2610010, DOE Genesis Program, Thinking Machines Lab and Cisco Research. This research used resources of the Oak Ridge Leadership Computing Facility, which is a DOE Office of Science User Facility supported under Contract DE-AC05-00OR22725.

\bibliographystyle{IEEEtran}
\bibliography{references}

\end{document}